\documentclass{agujournal}
\journalname{Geophysical Research Letters}

\begin{document}

%
%

\title{A hierarchical statistical framework for emergent constraints: application to snow-albedo feedback}

%
%

 \authors{Kevin W. Bowman \affil{1},
 Noel Cressie\affil{2}, Xin Qu\affil{3},
 and Alex Hall\affil{3}}

\affiliation{1}{Jet Propulsion Laboratory,
California Institute of Technology, Pasadena, California, USA}
\affiliation{2}{National Institute for Applied Statistics Research Australia (NIASRA), University of Wollongong, Australia}
\affiliation{3}{Department of Atmospheric and Oceanic Sciences, University of
California, Los Angeles, California, USA.}


\correspondingauthor{K.W.Bowman}{kevin.bowman@jpl.nasa.gov \copyright 2018. All rights reserved.}


\begin{keypoints}
\item A hierarchical   emergent constraints (HEC) framework  for climate projections is introduced.  
\item HEC depends on the signal-to-noise ratio between observational and climate uncertainty. 
\item Using HEC, the snow-albedo-feedback prediction interval  is found to be $(-1.25,-0.58)$  \%$K^{-1}$.
\end{keypoints}

%
%

\begin{abstract}
Emergent constraints use relationships between   future and current climate states to constrain projections of climate response.  Here, we introduce a statistical, hierarchical emergent constraint (HEC) framework in order to link future and current climate with observations.  Under Gaussian assumptions, the mean and variance of the future state is shown analytically  to be a function of the signal-to-noise (SNR) ratio between data-model error and current-climate uncertainty, and the correlation between future and current climate states.  We apply the HEC to the climate-change, snow-albedo feedback, which is related to the seasonal cycle in the Northern Hemisphere.   We obtain a snow-albedo-feedback prediction interval of $(-1.25, -0.58)$   \%$K^{-1}$.  The critical dependence on SNR and correlation  shows that neglecting these terms can lead to bias and  under-estimated uncertainty in constrained projections.   The flexibility of using  HEC under general assumptions throughout the Earth System is discussed.  
\end{abstract}

%
%

%



%
%
%

\section{Introduction}
\label{sec:intro}
The confrontation of  predictions with observations as a means of testing theories is a key demarcation of science and  critical to the advancement of scientific knowledge~\citep{Popper:1959aa}. In fields such as numerical weather prediction, data assimilation techniques provide a mathematical framework for narrowing the range or uncertainty of predictions through repeated evaluation against a broad suite of observations \citep{Tarantola:2006aa,Lewis:2006uq}.    Reducing the uncertainty in climate projections has been one of the signature challenges in Earth science.  In contrast to weather forecasting, the time  scales of climate projections do not permit ready validation. While historic and current observations can be used to benchmark climate models~\citep[e.g.,][]{Gleckler:2008hl,Teixeira:2014aa}, the establishment of robust relationships between contemporary performance and the credibility in future response has proven difficult.    One of the primary techniques  to explore these relationships is through climate-model ensembles~\citep{Collins:2007ys}. These ensembles may be derived from a core model where ``parametric" uncertainty is explored using, \emph{inter alia},    perturbed physics ensemble experiments~\citep[e.g.,][]{Allen:2000aa,Murphy:2004fk}.  Other approaches exploit ensembles from the Coupled Model Intercomparison Project (CMIP)~\citep{Taylor:2011aa,Eyring:2016aa} or similar MIPs in order to represent ``structural" uncertainty~\citep{Yokohata:2013aa}, which is the result of different  physical representations of processes.  These are used in weighting schemes that aim to provide the  ``best" combination of models rather than a strict model democracy~\citep[e.g.,][]{Tebaldi:2005ek,Smith:2009aa}.      The ``emergent" or ``observational" constraint approach,  as a means of using observations to indirectly reduce the uncertainty in climate projections, has only recently been appreciated \citep[e.g.,][]{Collins:2007ys,Collins:2012nx,Klein:2015aa,Cox:2018aa}.  

Here, an emergent constraint (EC)  is composed of 
\begin{enumerate}
	\item A dependence between  future climate, $z_{t+\tau}$, and current climate, $x_{t}$.
	\item A dependence between observations,  $y_{t}$, and current climate, $x_{t}$.  
\end{enumerate} 
Here, these dependencies will be expressed in terms of correlation. It is the synthesis of these quantities that yields an EC. For simplicity,  ``current" climate also refers to historic  climate.  Generally, a regression  between future  climate and  current climate  is  calculated empirically from a climate-model ensemble.  Through this relationship, model projections correlated with current-climate simulations  but inconsistent with observations should be treated with additional caution. These inconsistencies can signal where  more focused research is warranted \citep[e.g.,][]{DeAngelis:2015wd}.     

The EC approach has been applied to regional and global climate studies \citep[e.g.,][]{Hall:2006zg,Fasullo:2012fk,Qu:2014aa,Sherwood:2014zr,Borodina:2017aa,Cox:2018aa} and more broadly for Earth System studies \citep[e.g.,][]{Cox:2013uq,Bowman:2013fk,Wenzel:2016fk}.  These studies compute correlations between  $z_{t+\tau}$ and $x_{t}$ where they identify the range of models whose $x_{t}$ are within the precision of the observations, $y_{t}$.  However, they do not combine these factors to compute an estimate of future climate. For example,  \cite{Fasullo:2012fk} show that the relative humidity (RH) in the dry descending branch of circulation (300-500 hPa at 15$^{\circ}$) of most climate models is biased high in RH with respect to observations.  Models with high RH also tend to have lower equilibrium climate sensitivity (ECS). However, this study did not provide a quantitative method to incorporate the  present-future climate correlation ($-0.81$ in their case), the bias between observations and ensemble mean, and the observation uncertainty into an  estimate of ECS, thereby limiting the study to qualitative conclusions.  \cite{Cox:2018aa} provide a quantitative, probabilistic framework to estimate ECS given observations and current climate, however their   formulation does not include an explicit description of observational uncertainty.  This manuscript is careful to distinguish between $y_t$ and $x_t$ providing a framework that explicitly incorporates each of these  critical elements of EC. 

Here, we approach ECs from a  hierarchical statistical modeling  perspective \citep[e.g.,][]{Cressie:2011vn}.  The relationship between observations, states or processes, and parameters, are related through conditional probability distributions. Bayes' Theorem is employed to obtain a predictive distribution for these states  given the observations.  This framework allows us to give a prescriptive approach that integrates both model and observational uncertainties.  Moreover, it  has inherent recursive properties through  conditional distributions that subsume data assimilation algorithms such as  Kalman filtering, which are  implemented  in various forms in numerical weather prediction \citep[e.g.][]{Kalman:1960ke,Navon:2009nr,Wikle:2007fk}.  This approach has been applied to climate analysis, including regional-climate prediction and climate-change detection and attribution \citep{Kang:2013aa,Katzfuss:2017aa}.  

The hierarchical emergent constraint (HEC) framework introduced here explicitly relates future climate, current climate,   and observations through conditional-probability distributions that allow us to generalize previous EC studies.  For the purpose of illustration, we assume all distributions are Gaussian and  develop analytical formulas for these conditional distributions. This simplification allows for a  direct comparison to the EC literature where the Gaussian assumption is  implicit.  We then illustrate our approach by comparing the HEC to the ``classic" EC (CEC) for the snow-albedo feedback  of \cite{Hall:2006zg},   updated to use Climate Model Intercomparison Project-5 (CMIP5) models \citep{Taylor:2011aa,Qu:2014aa}.   The physical processes relating the current seasonal cycle and the future snow-albedo feedback are fairly straightforward, leading to  a causal interpretation of the correlative relationship between future and current climate ~\citep{Klein:2015aa}.   The HEC is subsequently used to explore how the correlation of the  future and current climate and the  observational uncertainty (expressed through a signal-to-noise ratio) impact  future climate-change estimates.   Challenges and future directions for this approach are discussed in the concluding section. 
%
%
%
%

%
%



\section{Methods}
\label{sec:meth}

\subsection{Hierarchical statistical framework for emergent constraints}
\label{sec:bayes}

A general probabilistic model of emergent constraints  is based upon the joint probability distribution of the future and  current climate given the observations, which can be written  as follows:
\begin{linenomath*}
	\begin{equation}
	[z_{t+\tau},x_{t}|y_{t}] \equiv \frac{[z_{t+\tau},x_{t},y_{t}]}{\int [z_{t+\tau},x_{t},y_{t}] dx_{t} dz_{t+\tau}},
		\label{eq:cond}
	\end{equation}
\end{linenomath*}
where 	
$z_{t+\tau}$, $x_{t}$, and $y_{t}$ are  random variables (or random vectors) representing the future climate, the current climate, and  the observations, respectively.  The bracket notation, $[x]$,  represents the probability density function of the variable $x$, and $[x|y]$ is the conditional probability density function of $x$ given $y$ \citep{Cressie:2011vn}.  The time indices, $t$ and $t+\tau$, ($\tau$>0) are used notationally to  distinguish between the current and future state. The term ``current" refers to both contemporary and historic states, or observations.  For simplicity,  these variables are referred to as states but can also be referred to as  processes (e.g., snow-albedo feedback).  The conditional density in Equation~\ref{eq:cond}  can be represented as:
\begin{linenomath*}
	\begin{equation}
	[z_{t+\tau},x_{t}|y_{t}]=[z_{t+\tau}|x_{t}][x_{t}|y_{t}],
	\label{eq:cond2}
	\end{equation}
\end{linenomath*}
since $[z_{t+\tau}|x_{t},y_{t}]=[z_{t+\tau}|x_{t}]$. That is, the conditional distribution of  future climate is  independent of  observations when the current climate is already known.   The future climate $z_{t+\tau}$ is predictable given $x_{t}$,  if $[z_{t+\tau}|x_{t}] \neq [z_{t+\tau}]$ \citep{DelSole:2007ta}.  Current climate $x_{t}$  is observable  if the observations $y_{t}$ satisfy $[x_{t}|y_{t}] \neq [x_{t}]$.  

A  hierarchical emergent constraint (HEC) can be defined as the probability of the future state  given observations of the current state; that is,  
\begin{linenomath*}
	\begin{equation}
	[z_{t+\tau}|y_{t}]= \int [z_{t+\tau}|x_{t}][x_{t}|y_{t}]dx_{t},
	\label{eq:zgy}
	\end{equation}
\end{linenomath*}
since $[z_{t+\tau}|y_{t}]$ is obtained as a marginal distribution of Equation~\ref{eq:cond}. An EC defined by Equation~\ref{eq:zgy} leads to a slightly different interpretation than the operational definition used in \cite{Cox:2018aa}, which is generally focused on $[z_{t+\tau}|x_{t}]$.  This difference can be critical when the data-model error on $y_{t}$ given $x_{t}$ is large.  That is, $[y_{t}|x_{t}]$ needs to be accounted for, as does knowledge of $x_{t}$, leading to an appropriate $[x_{t}|y_{t}]$ on the right-hand side of Equation~\ref{eq:zgy}.   As we demonstrate below, a weak correlation between $z_{t+\tau}$ and $x_{t}$ can be partially offset by a strong correlation  between $y_{t}$ and $x_{t}$ (i.e., how well the climate is observed), and vice versa.  Consequently, the predictability and observability of climate are inextricably linked.  

The inference of $x_{t}$ from observations, $y_{t}$, frequently use Bayesian techniques \citep[e.g.,][]{Rodgers:imastp-00}:
\begin{linenomath*}
	\begin{equation}
	[x_{t}|y_{t}] =  \frac{[y_{t}|x_{t}][x_{t}]}{[y_{t}]}.
	\label{eq:bayes}
	\end{equation}
\end{linenomath*}
As will be shown in the next section, $[x_{t}|y_{t}]$ accounts for the uncertainty of the observing system (data-model errors) and the uncertainty of the state (state error). Substitution of Equation~\ref{eq:bayes} into Equation~\ref{eq:zgy} leads to 
\begin{linenomath*}
	\begin{equation}
	[z_{t+\tau}|y_{t}]=\frac{1}{[y_{t}]} \int [z_{t+\tau}|x_{t}][y_{t}|x_{t}][x_{t}]dx_{t}.
	\label{eq:zgybayes}
	\end{equation}
\end{linenomath*}
The distributions inside the integral of Equation~\ref{eq:zgybayes} are typically straightforward, but the denominator, $y_{t}$,  can be problematic to compute. However, when Gaussian assumptions are made, it has an analytical form and can be evaluated easily.  
\subsection{Application to linear Gaussian constraints}
 \label{sec:gaussian}
ECs in the literature  frequently express the relationship between the future and current state in terms of a simple linear regression. These regressions can in turn be interpreted as the first moment of a conditional density of Gaussian distributions.  In the leading case  where $z_{t+\tau}$, $x_{t}$, and $y_{t}$ are jointly Gaussian,  a closed-form expression for [$z_{t+\tau}|y_{t}$] can be obtained analytically.  

Assume that the observations are related to current climate through
\begin{linenomath*}
	\begin{equation}
	y_{t}= x_{t} + n_{t},
	\label{eq:addmod}
	\end{equation}
\end{linenomath*}
where $x_{t}\sim \mathcal{N}(\mu_{x_{t}},\sigma^{2}_{x_{t}})$ and  $n_{t}\sim \mathcal{N}(0,\sigma^{2}_{n_{t}})$  are  independent Gaussian random variables parameterized by their mean and variance. The observation, $y_{t}$, is a measurement of the true climate state, $x_{t}$, and  the error, $n_{t}$, which incorporates multiple sources of uncertainty including systematic and random error from the measurement along with representation and model error \citep{Brasseur:2017aa}. There may be bias in the error, which would lead to a non-zero mean. If the bias can be determined, then it can be subtracted from the observation, $y_{t}$, in Equation~\ref{eq:addmod}. 

Combining Equation~\ref{eq:addmod} with Equation~\ref{eq:bayes}, we obtain 
\begin{linenomath*}
	\begin{equation}
	[y_{t}|x_{t}][x_{t}] \propto \exp\left[-\frac{(y_{t}-x_{t})^{2}}{2\sigma^{2}_{n_{t}}}\right] \exp\left[-\frac{(x_{t}-\mu_{x_{t}})^{2}}{2\sigma^{2}_{x_{t}}}\right].
	\label{eq:gauss}
	\end{equation}
\end{linenomath*}
 Then, the Maximum A Posteriori (MAP) estimate  of $x_{t}$ is the  expectation of the conditional  distribution in Equation~\ref{eq:bayes}, which can be calculated from Equation~\ref{eq:gauss} to be
\begin{linenomath*}
	\begin{equation}
	   \label{eq:exgy}	E(x_{t}|y_{t}) = \mu_{x_{t}}+K_{x_{t},y_{t}}(y_{t}-\mu_{x_{t}}),
	\end{equation}
and  the conditional variance is
    \begin{equation}
        \label{eq:varxgy}	var(x_{t}|y_{t})  =  (1-K_{x_{t},y_{t}})\sigma^{2}_{x_{t}},
    \end{equation}
\end{linenomath*}
where 
\begin{linenomath*}
	\begin{equation}
	K_{x_{t},y_{t}} = \frac{\sigma^{2}_{x_{t}}}{\sigma^{2}_{x_{t}}+\sigma^{2}_{n_{t}}}.
	\label{eq:gain}
	\end{equation}
\end{linenomath*}
The quantity $K_{x_{t},y_{t}}$ is the ``gain" of the MAP estimator, which  balances the uncertainty in the current state with the precision of the observation~\citep{Wikle:2007fk}. 

Under our assumptions,  $z_{t+\tau}$ and $x_{t}$ are jointly Gaussian. Consequently,  the future climate state conditioned on the current state is 
\begin{linenomath*}
	\begin{eqnarray}
\label{eq:ezgx}	E(z_{t+\tau}|x_{t}) =\mu_{z_{t+\tau}|x_{t}} &=& \mu_{z_{t+\tau}}+K_{z_{t+\tau},x_{t}}(x_{t}-\mu_{x_{t}})\\ 
\label{eq:varzgx} var(z_{t+\tau}|x_{t}) = \sigma^2_{z_{t+\tau}|x_{t}}  & = & \sigma^{2}_{z_{t+\tau}} -K^{2}_{z_{t+\tau},x_{t}}\sigma^{2}_{x_{t}}  = (1-\rho^2)\sigma^{2}_{z_{t+\tau}}, 
	\end{eqnarray}
\end{linenomath*}
where 
\begin{linenomath*}
	\begin{equation}
	K_{z_{t+\tau},x_{t}} = \frac{\rho \sigma_{z_{t+\tau}}}{\sigma_{x_{t}}},
	\end{equation}
\end{linenomath*}
and $\rho$ is the correlation coefficient between $z_{t+\tau}$ and $x_{t}$.  The expectation in Equation~\ref{eq:ezgx} can be cast as a simple ``straight-line" approximation between these states, namely, $z_{t+\tau} = \alpha + \beta x_{t}$, where $\beta = K_{z_{t+\tau},x_{t}}$ is the slope, and  $\alpha = \mu_{z_{t+\tau}}-K_{z_{t+\tau},x_{t}}\mu_{x_{t}}$ is the intercept.  It  is this "straight-line" approximation in ECs that is fitted from climate-model ensembles.    

Equations \ref{eq:exgy}--\ref{eq:varxgy} and \ref{eq:ezgx}--\ref{eq:varzgx} define up-to-second-order statistical descriptions of   the current state given observations, (i.e., $[x_{t}|y_{t}]$) and the future state given the current state, (i.e., $[z_{t+\tau}|x_{t}]$).  Under Gaussian assumptions, all conditional densities $[z_{t+\tau}|x_{t}]$, $[x_{t}|y_{t}]$, and $[z_{t+\tau}|y_{t}]$ are Gaussian, and hence only their first and second moments are needed. 

The derivations that follow up to Equation~\ref{eq:normvar}, do not require Gaussian assumptions.  Applying the  law of iterated expectations \citep{Ross:2010aa},  the first moment  of $[z_{t+\tau}|y_{t}]$  is
\begin{linenomath*}
	\begin{equation}
	E(z_{t+\tau}|y_{t}) =  E(E(z_{t+\tau}|y_{t},x_{t})) = E(E(z_{t+\tau}|x_{t})|y_{t}).
	\label{eq:lawitexp}	
	\end{equation}
\end{linenomath*}
Substituting Equation~\ref{eq:ezgx} into Equation~\ref{eq:lawitexp}, yields
\begin{linenomath*}
	\begin{equation}
	E(z_{t+\tau}|y_{t}) =   \mu_{z_{t+\tau}} + K_{z_{t+\tau},x_{t}} (E(x_{t}|y_{t})-\mu_{x_{t}}).
		\label{eq:zgyv1}
	\end{equation}	
\end{linenomath*}
Equation~\ref{eq:exgy} can then be substituted into Equation~\ref{eq:zgyv1}, yielding:
\begin{linenomath*}
	\begin{eqnarray}
	E(z_{t+\tau}|y_{t}) =\mu_{z_{t+\tau}|y_{t}} & = & \mu_{z_{t+\tau}} + K_{z_{t+\tau},x_{t}}K_{x_{t},y_{t}}(y_{t}-\mu_{x_{t}}) \\ 
	 & = & \mu_{z_{t+\tau}} + \frac{\rho \sigma_{z_{t+\tau}}\sigma_{x_{t}}}{\sigma^{2}_{x_{t}}+\sigma^{2}_{n_{t}}}(y_{t}-\mu_{x_{t}}). \label{eq:ezgy}
	\end{eqnarray}
\end{linenomath*}
Defining the statistically normalized anomaly of the  future-climate estimate,  $ \delta_{z_{t+\tau}|y_{t}} \equiv (\mu_{z_{t+\tau}|y_{t}}-\mu_{z_{t+\tau}})/ \sigma_{z_{t+\tau}} $ and the  normalized anomaly of the current-climate estimate, $ \delta_{x_{t}|y_{t}} \equiv (y_{t}-\mu_{x_{t}})/\sigma_{x_{t}}$, Equation~\ref{eq:ezgy} can be rearranged to yield
\begin{equation}\label{eq:normanom}
 \frac{\delta_{z_{t+\tau}|y_{t}}}{\delta_{x_{t}|y_{t}}} = \frac{\rho}{1+(\sigma^{2}_{x_{t}}/\sigma^{2}_{n_{t}})^{-1}}.
\end{equation}
The magnitude of the normalized update is driven by  $\rho$ and $(\sigma^{2}_{x_{t}}/\sigma^{2}_{n_{t}})$.  The first term is the correlation between the future and current climate, and the second quantity is
\begin{equation}\label{eq:snr} 
\mathrm{SNR} \equiv \frac{\sigma^{2}_{x_{t}}}{\sigma^{2}_{n_{t}}},
\end{equation}
which defines the relative strength of the signal variability  to the noise variability. If the signal dominates the noise, then the update of the anomaly ratio in Equation~\ref{eq:normanom} is controlled by  $\rho$. Conversely, if noise dominates then, as expected, the forecast anomaly will be close to zero.  Notice that the normalized anomaly in the future climate estimate  $z_{t+\tau}$ is proportional to $\rho$. For $\rho=1$, the update in Equation~\ref{eq:normanom} is controlled by the SNR and, for $\rho=0$, no update is zero.  

 In order to calculate the variance (the central second moment of $[z_{t}|y_{t}]$) of the EC,  the law of total variance \citep[e.g.,][]{Ross:2010aa} is invoked: 
\begin{linenomath*}
	\begin{equation}
	var(z_{t+\tau}|y_{t}) =  E(var(z_{t+\tau}|x_{t})|y_{t})+var(E(z_{t+\tau}|x_{t})|y_{t}).  
	\label{eq:lawtotvar}
	\end{equation}
\end{linenomath*}
By substituting Equations~\ref{eq:varxgy},~\ref{eq:ezgx}, and~\ref{eq:varzgx} into the right-hand side of  Equation~\ref{eq:lawtotvar},  the variance of the HEC can be written as 
\begin{linenomath*}
	\begin{eqnarray}
	var(z_{t+\tau}|y_{t}) = \sigma^{2}_{z_{t+\tau}|y_{t}} & = & var(z_{t+\tau}|x_{t}) + K^{2}_{z_{t+\tau},x_{t}}var(x_{t}|y_{t}) \\ 
	& = & \sigma^{2}_{z_{t+\tau}}-K^{2}_{z_{t+\tau},x_{t}} K_{x_{t},y_{t}}\sigma^{2}_{x_{t}} \\ 
	& = & \left(1-\frac{\rho^{2}}{1+(\sigma^{2}_{x_{t}}/\sigma^{2}_{n_{t}})^{-1}}\right)\sigma^{2}_{z_{t+\tau}} \label{eq:varzgy}
	\end{eqnarray}
\end{linenomath*}
The right-hand side of Equation~\ref{eq:varzgy} can be normalized to compute a relative reduction in variance:
\begin{equation}\label{eq:normvar}
\frac{\sigma^{2}_{z_{t+\tau}|y_{t}} }{\sigma^{2}_{z_{t+\tau}}} = \left(1-\frac{\rho^{2}}{1+(\sigma^{2}_{x_{t}}/\sigma^{2}_{n_{t}})^{-1}}\right),
\end{equation}
which is always between 0 and 1.

Under Gaussian assumptions,  
Equations~\ref{eq:ezgy} and~\ref{eq:varzgy} provide a complete description of the dependence of the future climate's distribution given the observations, $y_{t}$, of the current climate.  There are some important limiting conditions that illuminate the relationships between future climate, current climate, and observations. Similar to Equation~\ref{eq:normanom}, the change in uncertainty is driven by the interplay between $\rho$ and the SNR.    As  $\rho\rightarrow 0$, then $[z_{t+\tau}|y_{t}]$ converges in distribution to $[z_{t+\tau}]$, which is expected. That is, if observations are uncorrelated with the future state, they will have no impact on the uncertainty of that state.  If the SNR is high (i.e., $\sigma^2_{x_t} >> \sigma^2_{n_{t}}$), then the relative reduction in the uncertainty of the future state is controlled completely by the correlation through $1-\rho^2$, and the CEC coincides with the HEC we have developed here. 	
\section{Application to snow-albedo feedback}

\label{sec:res}
 The snow-albedo feedback (SAF) is an important component of the global hydrological response to increases in carbon dioxide \citep{Bony:2006vn}. The SAF accounts for much of the spread in Northern Hemispheric (NH) landmass warming \citep{Qu:2014aa}.  This feedback amplifies global and NH warming since  snow-cover retreat accelerates when mean temperatures increase, which exposes a lower albedo surface that in turn  leads to a decrease in net shortwave (SW) radiation, $\bar{Q}_{\mathrm{net}}$, at the top-of-the-atmosphere (TOA).  The SAF-induced change in $\bar{Q}_{\mathrm{net}}$ to surface temperature  can be expressed as 
 \begin{equation}\label{eq:qnet}
 \frac{\Delta \bar{Q}_{\mathrm{net}}}{\Delta \bar{T}_{s}} = -\frac{1}{A_{R}} \int_{S,R} Q(t,r) \frac{\partial\alpha_{p}}{\partial\alpha_{s}}(t,r) \frac{\Delta\alpha_{s}}{\Delta\bar{T}_{s}}(t) dt dr,
 \end{equation}
 where  $Q(t,r)$ is the incoming radiation at TOA at time $t$ and position $r$, $\alpha_{p}$ is the planetary albedo, $\alpha_{s}$ is the surface-albedo, $A_{R}$ is the
 area of region $R$,  $\Delta\bar{T}_{s}$  is a change in the regionally averaged surface air temperature, and the domains of integration are  the annual cycle, $S$, and a region $R$.  Annually, $\Delta \bar{Q}_{\mathrm{net}}$  is controlled in part by the magnitude of  the  snow-albedo  sensitivity to temperature, which is $\Delta\alpha_{s}(t)/\Delta \bar{T}_{s}(t)$, especially  during the NH Spring.  The sensitivity, which is a function of snow cover and vegetation type,  is climate-model dependent.  \cite{Qu:2014aa} showed that there is a strong correlation between climate models with a large sensitivity  and their change in NH landmass temperatures through the snow-albedo feedback. 
 
 In order to diagnose the snow-albedo feedback in climate models, $\Delta \overline{\alpha}^{sc}_s$ and  $\Delta \overline{T}^{sc}_s$ are  defined as the difference between April and May surface albedo and temperature, respectively,  averaged over  NH extratropical landmasses.  Their ratio, $\Delta\overline{\alpha}^{sc}_{s}/\Delta \overline{T}^{sc}_{s}$, is the  seasonal cycle snow-albedo temperature sensitivity (SCSAT).  Likewise,  $\overline{\alpha}^{cc}_s$ and $\Delta \overline{T}^{cc}_s$   are quantified by the difference in  April values between the current (1980--1999) and future (2080--2099) climates regionally averaged over NH extratropical landmasses. Their ratio is the  climate-change snow-albedo temperature sensitivity (CCSAT).
 
\begin{figure}[ht!]
	\centerline{\includegraphics[width=0.7\linewidth]{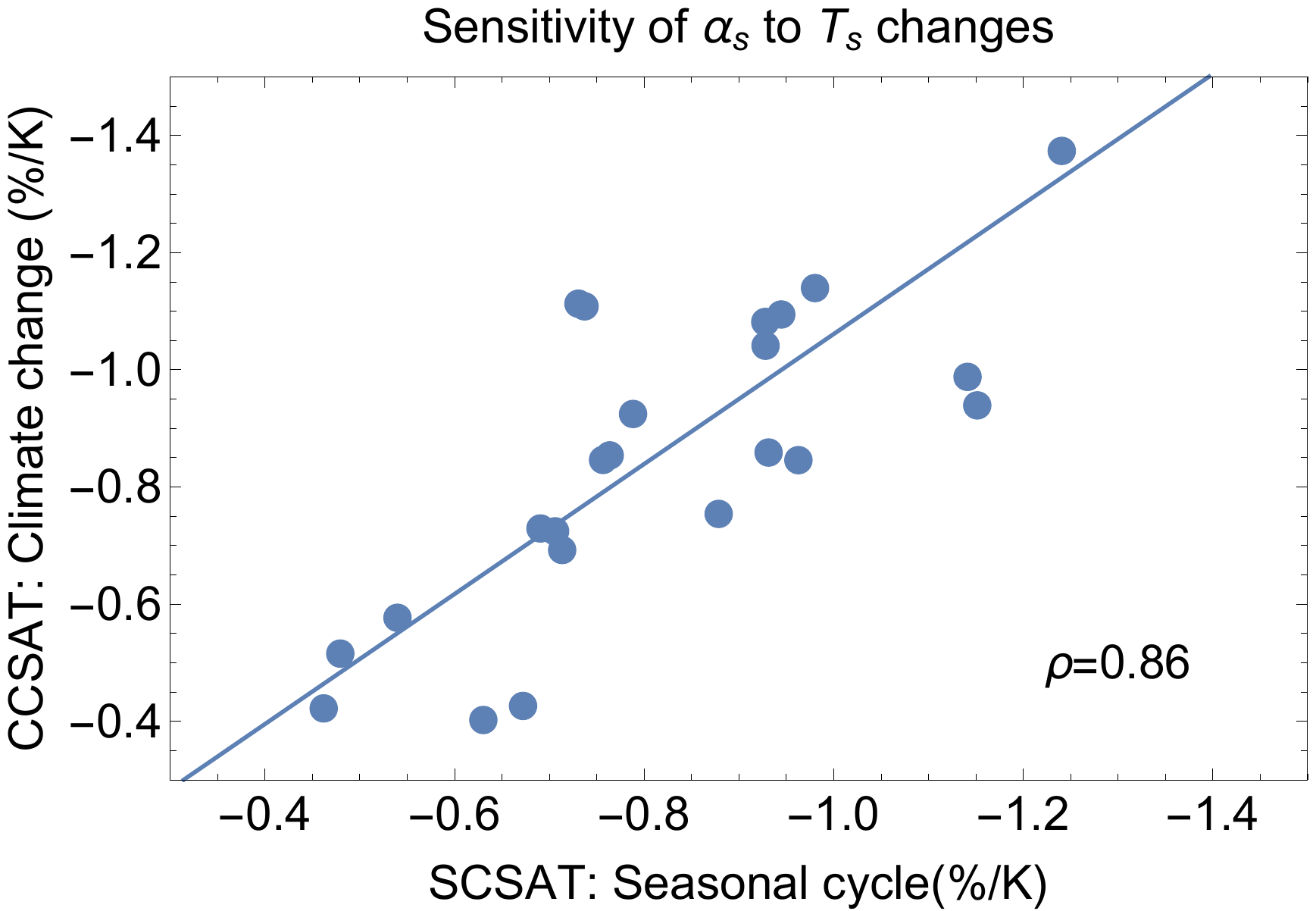}}
	\caption{Linear regression of the climate-change snow-albedo temperature (CCSAT) against the seasonal cycle snow-albedo temperature sensitivity (SCSAT)   defined over Northern Hemispheric land.  Each of the 25 dots represents a CMIP5 model computed in \protect{\cite{Qu:2014aa}}.    }
	\label{fig:gsar5fabr5}
\end{figure}

 Figure~\ref{fig:gsar5fabr5} shows the regression between   CCSAT and  SCSAT across  25 CMIP5 models.  The regression is characterized by a strong correlation ($\rho$ = 0.86) along with a slope (1.11)  and intercept (0.05 $\%K^{-1}$) that are close to unity and zero, respectively.  The  correlation here reflects  a physical relationship defined through Equation~\ref{eq:qnet}.  Consequently, the SAF in the contexts of both seasonal cycle (current) and climate change (future) is similarly influenced by these physically related processes \citep{Qu:2014aa}.    
 
\begin{figure}
	\centering
	\includegraphics[width=0.7\linewidth]{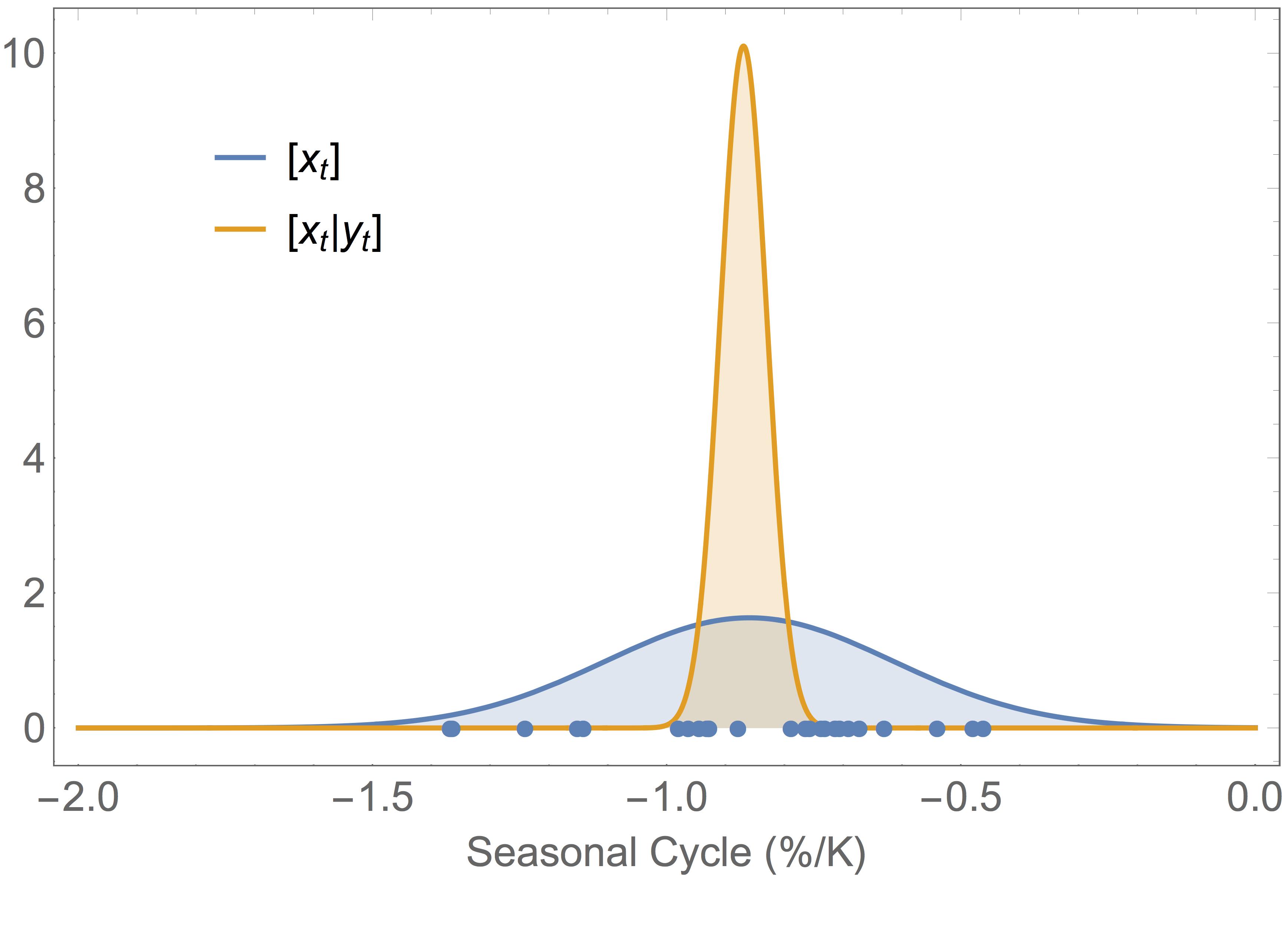}
	\caption{ The fitted Gaussian probability density functions (pdf) of seasonal cycle snow-albedo temperature sensitivity (SCSAT), which is defined as the ratio of the change in snow-albedo between May and June normalized by the change in temperature.   Individual model-based estimates are represented by dots along the abscissa. }
	\label{fig:gssar5gobsv2}
\end{figure}

 Applying the HEC framework to SAF, we define $x_{t}= \Delta\overline{\alpha}^{sc}_{s}/\Delta\overline{T}^{sc}_{s}$, which  is  SCSAT and $z_{t+\tau} = \overline{\alpha}^{cc}_s/\Delta \overline{T}^{cc}_s$, which is  CCSAT.     Figure~\ref{fig:gssar5gobsv2} shows the spread of the SCSAT as a Gaussian probability density function   estimated from  the CMIP5 models and is denoted as   $[x_{t}]$.  \cite{Qu:2014aa} showed that the model-spread in the seasonal cycle is attributable to the mean effective snow albedo, which in turn is controlled primarily by land-surface modeling (e.g., vegetation canopy).   The  CMIP5 SCSAT first-order and (square root) second-order moments are $\mu_{x} = -0.860~\%K^{-1}$ and $\sigma_{x} = 0.244~\%K^{-1}$, respectively. Consequently  the variability is about 30\% of the absolute mean.  
 
 An observational constraint  on  SCSAT is based upon a combination of MODIS surface-albedo measurements from 2001--2012 \citep{Jin:2003aa} and surface air temperature from ERA-interim \citep{Dee:2011ys}. We refer the interested reader to \cite{Qu:2014aa} for additional details.  The observed  SCSAT is $y_{t} =-0.87 $ \%K$^{-1}$ with an observational uncertainty of $\sigma_{n_{t}} = 0.04$ \%K$^{-1}$.  Consequently, the one-standard-deviation range is $[-0.92,-0.83]$ \%K$^{-1}$. Figure~\ref{fig:gssar5gobsv2} shows the observationally constrained distribution calculated from Equations~\ref{eq:exgy} and \ref{eq:varxgy}.
The conditional mean  is $\mu_{x_{t}|y_{t}}=-0.870$ \%K$^{-1}$, and the conditional uncertainty is $\sigma_{x_{t}|y_{t}}=0.0395$ \%K$^{-1}$. Hence, $\sigma_{x_{t}|y_{t}}$ is about 6 times less than $\sigma_{x_{t}}$,the uncertainty in the  unconditional distribution of the state $x_{t}$. 
 
\begin{figure}
	\centering
	\includegraphics[width=0.7\linewidth]{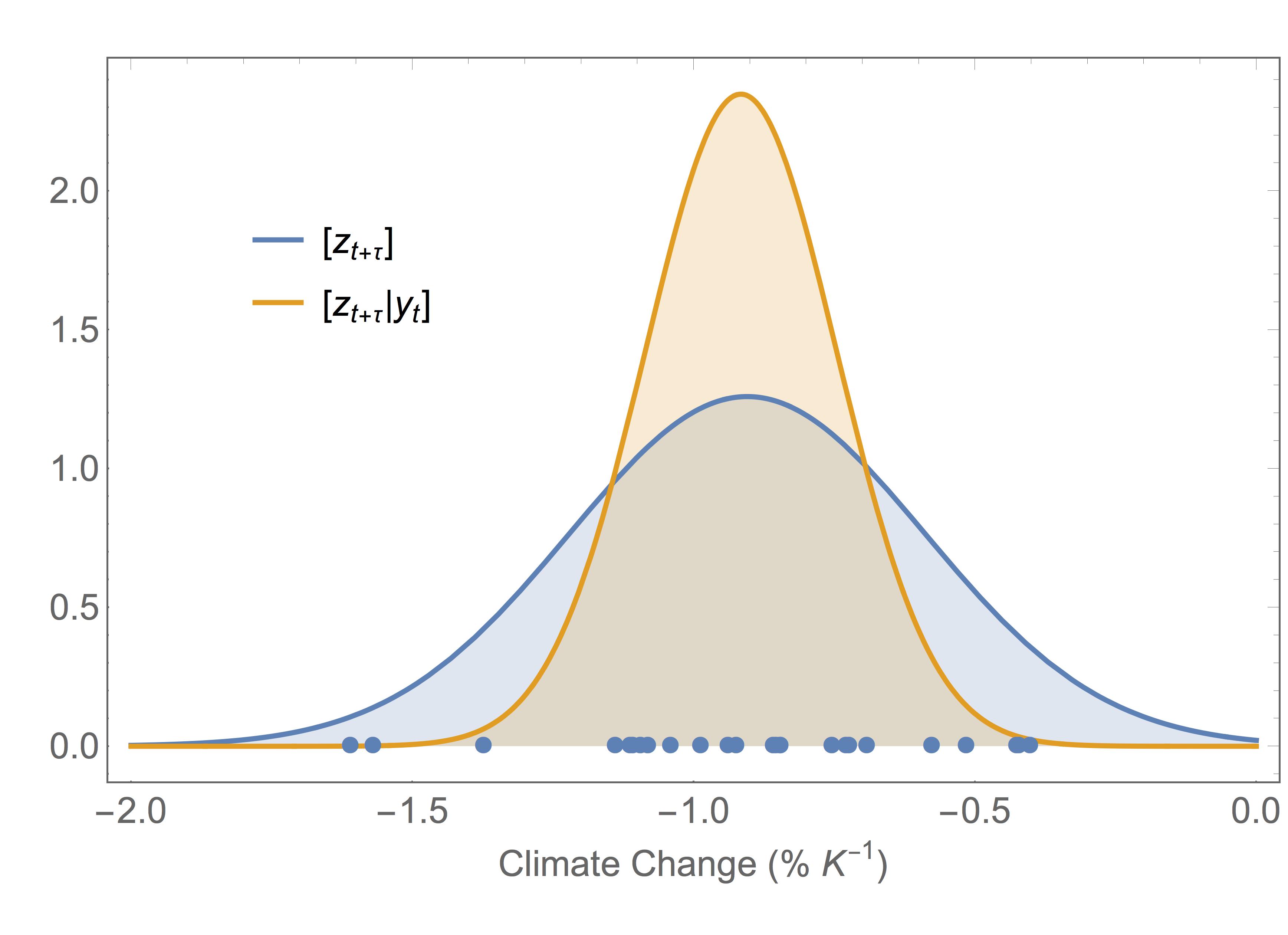}
	\caption[emergent constraint]{ The fitted Gaussian probability density functions (pdf) of climate-change snow-albedo temperature sensitivity (CCSAT), which is defined as the ratio of the change in snow-albedo in April between present day (1980-1999) and future (2080-2099) normalized by the change in regionally averaged NH land  temperatures averaged over the same time periods.   Individual model estimates of the CCSAT are represented by dots along the abscissa.}
	\label{fig:gpar5gobs}
\end{figure}

The effects of  observations on the predictive distribution of CCSAT is the critical question, which was not addressed in  \cite{Qu:2014aa}.  It would be tempting to simply use the regression line itself to project the observational estimate of SCSAT into CCSAT.   However, that would neglect the critical role of $\rho$ and SNR.  The HEC of CCSAT, $[z_{t+\tau}|y_{t}]$, is shown in Figure~\ref{fig:gpar5gobs} and computed from Equation~\ref{eq:zgy} assuming jointly Gaussian distributions with a conditional mean (computed from Equation~\ref{eq:ezgy}) to be $\mu_{z_{t+\tau}|y_{t}}=-0.916$  \%$K^{-1}$, 
and the conditional  uncertainty (computed from Equation~\ref{eq:varzgy}) to be  
$\sigma_{z_{t+\tau}|y_{t} }= 0.170$ \%$K^{-1}$.  
The conditional mean is similar to the mean of the state,  $\mu_{z_{t+\tau}} = -0.905$ \%K$^{-1}$, 
but the conditional uncertainty, $\sigma_{z_{t+\tau}|y_{t} }$ is about 1.9 times less than $\sigma_{z_{t+\tau}}$ =  0.317 \%K$^{-1}$. This is seen from Equation~\ref{eq:normvar} where  the role of $\rho=0.86$ and SNR=$(6.1)^2$ is crucial here.  

 The HEC can now be used to obtain the predictive distribution  of CCSAT given observations, which goes beyond the results in \cite{Hall:2006zg} and \cite{Qu:2014aa}.  Based on the marginal distribution $[x_{t}]$ from the CMIP5 models, the 95\% prediction interval of CCSAT is $(-1.425,-0.385)$~\%$K^{-1 }$. With HEC, the 95\% prediction interval has narrowed substantially to $(-1.25,-0.58)$~\%$K^{-1}$. 
 
\section{Role of the signal-to-noise ratio and correlation in HEC}
\begin{figure}
	\centering
	\includegraphics[width=0.7\linewidth]{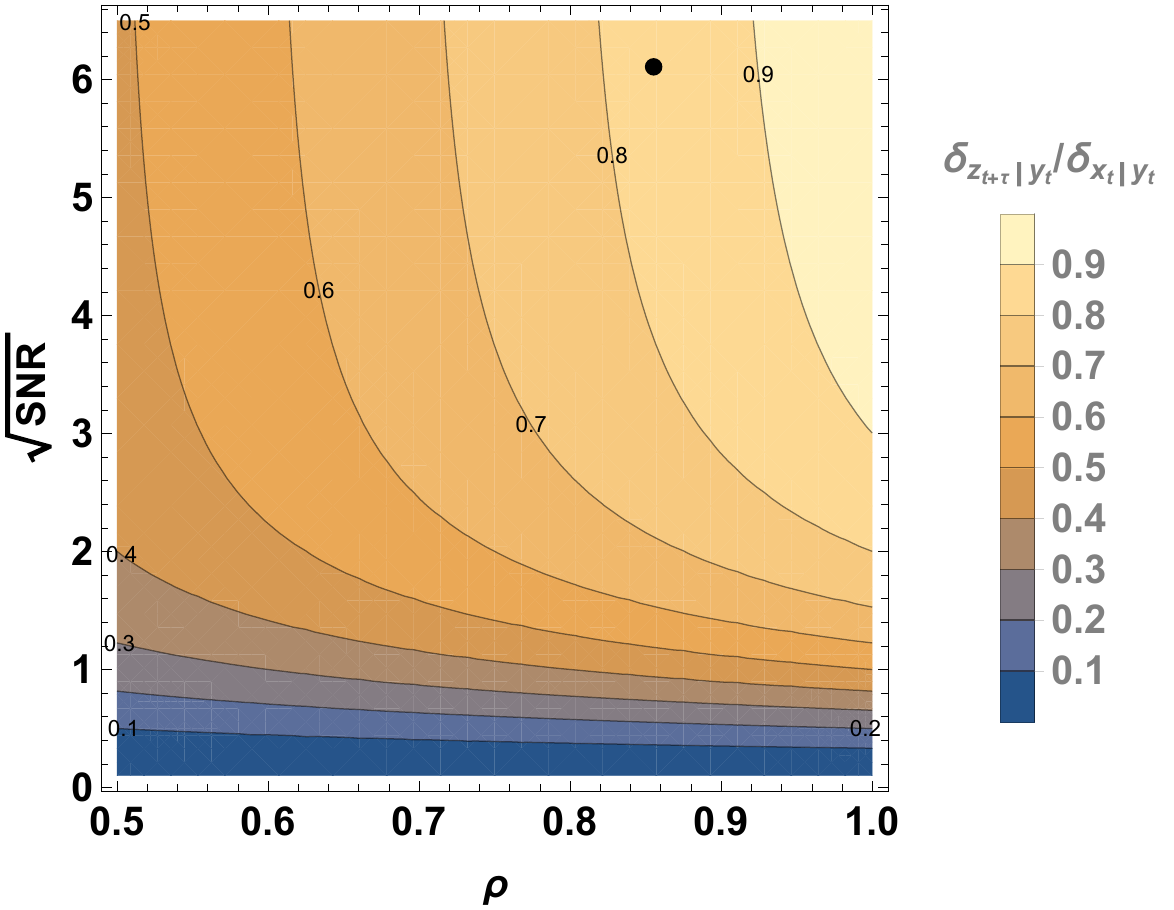}
	\caption[Mean deviation]{Normalized change in the predicted anomaly in the future state as a function of correlation ($\rho$) and signal-to-noise ratio (SNR). The black dot on the 0.83 contour  is positioned at ($\rho$, $\sqrt{\mathrm{SNR}}$)=(0.86, 6.1)  for the CMIP5 models and observations from \protect{\cite{Qu:2014aa}}.}
	\label{fig:gmeanar5gobssnr}
\end{figure}

\begin{figure}
	\centering
	\includegraphics[width=0.7\linewidth]{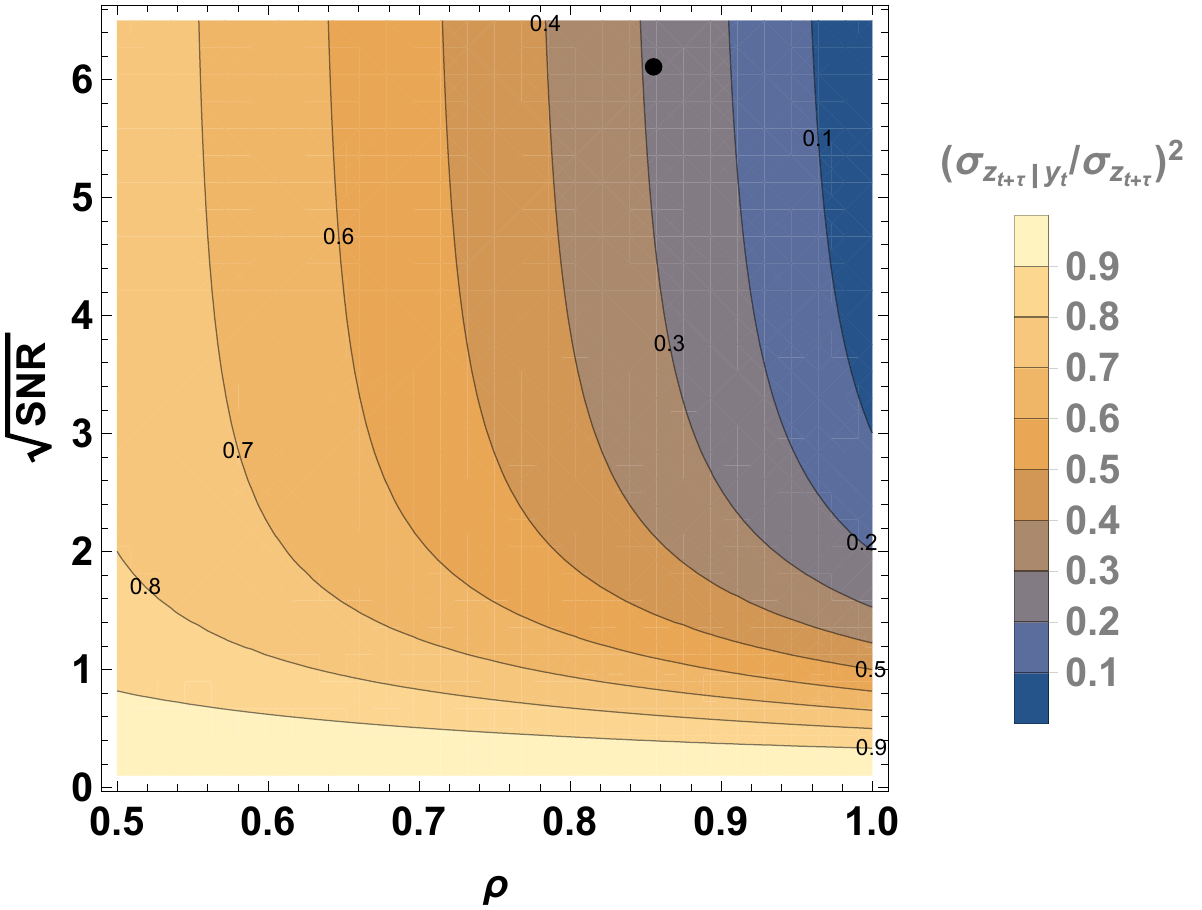}
	\caption[trade-off]{Normalized reduction factor as a function of correlation, $\rho$, and signal-to-noise ratio (SNR). The black dot on the 0.29 contour is the ($\rho$, $\sqrt{\mathrm{SNR}}$)=(0.86, 6.1) for the CMIP5 models and observations from \protect{\cite{Qu:2014aa}}.}
	\label{fig:gvarar5gobssnrv3}
	\end{figure}

The snow-albedo feedback provides a good starting point to consider HEC in a  broader context.  The analytical solution shows that in a normalized form (Equations \ref{eq:normanom} and~\ref{eq:normvar}), the conditional mean and the conditional variance in the update are  a  function of only the correlation $\rho$  and the SNR.   Increases in $\rho$ and SNR both act in the HEC to reduce the uncertainty in future climate.  To explore these two mechanisms, Figures~\ref{fig:gmeanar5gobssnr} and \ref{fig:gvarar5gobssnrv3}  show the normalized conditional mean update (Equation~\ref{eq:normanom})  and  the normalized  uncertainty reduction factor  (Equation~\ref{eq:normvar}) resulting from the use of the HEC. The black circles in those figures are computed based upon  $\rho$=0.86  and  $\sqrt{\mathrm{SNR}}=6.1$ obtained for the SAF study.  

 A one standard deviation anomaly for both the normalized CCSAT and  SCSAT  is unity.  The the right-hand side of Equation~\ref{eq:normanom} is the weight that scales the CCSAT anomaly ($\delta_{z_{t+\tau}|y_{t}}$) relative to the SCSAT anomaly ($\delta_{x_{t}|y_{t}}$). For a perfect correlation and high SNR, the weight itself is nearly unity; a normalized anomaly in CCSAT is equal to a normalized anomaly in SCSAT ($\delta_{z_{t+\tau}|y_{t}}/\delta_{x_{t}|y_{t}}\approx1$). The range of weights is shown in Figure~\ref{fig:gmeanar5gobssnr} where the black dot is on the 0.83 contour corresponding to ($\rho$,  $\sqrt{\mathrm{SNR}}$)=(0.86, 6.1) for this study.  Figure~\ref{fig:gvarar5gobssnrv3} shows the contour plot of  the  reduction factor in the variance  (i.e., $\sigma^{2}_{z_{t+\tau}|y_{t} }/\sigma^{2}_{z_{t+\tau}}$ from  Equation~\ref{eq:normvar})  for CCSAT; the black dot  is on the 0.29 contour again corresponding to ($\rho$, $\sqrt{\mathrm{SNR}}$)=(0.86, 6.1).   It is important to note that this reduction is not unique to the SAF study.  Any HEC with the same correlation and SNR will yield a variance reduction of approximately 30\%.  For a hypothetical HEC with the same correlation as the SAF,  but a much smaller SNR of 1, the normalized update would be reduced  to  $\rho/2$=0.43 (as compared to 0.83 in our case) and the variance reduction factor of $1-\rho^2/2=$ 0.8 (as compared to 0.54 in our case).   This interplay, especially the role of observational uncertainty, is not found in the formulation of \cite{Cox:2018aa}. We have shown that neglecting the role of observational uncertainty, which is the case in classic EC, can lead to incorrect prediction intervals especially for lower precision data  such as those discussed in \cite{Fasullo:2012fk} and \cite{Sherwood:2014zr}. 
\section{Conclusions}
Projections of change in the Earth System from anthropogenic forcing is one of the defining challenges in climate science.  ECs represent an important approach to incorporating observations  into climate-model projections that  relate present-day variability to future response.  In this work,  ECs are explicitly defined through Equation~\ref{eq:zgy} as conditional distributions  within a HEC framework.   Classical EC studies frequently use a linear regression  but do not  account  for both the  correlation  between $z_{t+\tau}$ and $x_{t}$ and  the precision in observing $x_{t}$ with $y_{t}$ as does  HEC. The formulation of  the Maximum A Posterior (MAP) solution in Equation~\ref{eq:ezgy} more directly links EC with data assimilation techniques.  For  non-Gaussian processes, more advanced tools such as Markov Chain Monte Carlo (MCMC) could be used to compute accurate prediction intervals based on the HEC \citep{Cressie:2011vn}.

Like any statistical approach,  assessing whether $[z_{t+\tau}|x_{t}]$ is  causal  remains an important challenge \citep{Klein:2015aa}.  While this work does not explicitly address these considerations, the HEC framework introduced here more readily allows EC to be linked to causality analysis \citep[e.g.][]{Pearl:2009aa,Sugihara:2012aa}.  

We note the  joint distribution $[z_{t+\tau},x_{t}]$ is dependent on the climate-model ensemble, which may not be robust to model choice or may systematically miss important processes. Increased and systematic use of observations and high-resolution modeling can improve confidence in Earth-System models \citep{Schneider:2017aa}, of which this approach can readily take advantage.  Current applications of EC have generally used averaged scalar processes $z_{t+\tau}$, $x_{t}$, and $y_{t}$.    Including multiple types of observations $y^{(1)}_{t}$, $y^{(2)}_{t}$, $\ldots$ sensitive to $x_{t}$ within this framework, will provide more information for implementing ECs from processes simulated in climate models. This is especially true for critical climate metrics such as the  equilibrium climate sensitivity, which depends on multiple processes  including water vapor, clouds, and snow-albedo feedbacks.  We would expect that incorporating multiple measurements   that are sensitive to a range of these key feedbacks will ultimately be necessary to constrain climate projections.  That extension is a subject of future research. 

\clearpage


%

\acknowledgments
KB's research was carried out at the Jet Propulsion Laboratory, California Institute of Technology, under a contract with the National Aeronautics and Space Administration. KB was supported under NASA ROSES NNH13ZDA001N-AURAST.  NC's research was supported by an ARC Discovery Project DP150104576. XQ and AH are supported by NSF Grant 1543268, titled ``Reducing Uncertainty Surrounding Climate Change Using Emergent Constraints."

%
%
%
%
%
%
%
%
%

%
%





\end{document}